\documentclass[twocolumn,showpacs,superscriptaddress,prl]{revtex4}
\usepackage{amsbsy,amssymb,amsmath,bm}
\usepackage{graphicx,color}

\newcommand{\av}[1]{\langle #1 \rangle}

\newcommand{\cP}{{\cal P}}

\renewcommand{\Im} {\mathop{\mathrm{Im}}}

\begin{document}
\title{Low-frequency noise in tunneling through a single spin}
\author{Y. M. Galperin}
\affiliation{Department of Physics, University of Oslo, PO Box
1048
  Blindern, 0316 Oslo, Norway}
\affiliation{A. F. Ioffe  Physico-Technical Institute of Russian
  Academy of Sciences, 194021
  St. Petersburg, Russia}
\affiliation{Argonne National Laboratory, 9700 S. Cass av.,
Argonne,
  IL 60439, USA}
\author{V. I. Kozub}
\affiliation{A. F. Ioffe  Physico-Technical Institute of Russian
  Academy of Sciences, 194021
  St. Petersburg, Russia}
\affiliation{Argonne National Laboratory, 9700 S. Cass av.,
Argonne,
  IL 60439, USA}
\author{V. M. Vinokur}
\affiliation{Argonne National Laboratory, 9700 S. Cass av.,
Argonne,
  IL 60439, USA}
\date{\today}
\begin{abstract}
We propose measurements of low-frequency noise in the tunneling
current through a single molecule with a spin as an experimental
probe for identifying a mechanism of the spin-dependent tunneling.
A specific tail near the zero frequency in the noise spectrum is
predicted; the amplitude and  the width of being of the same order
of magnitude as the recently reported peak in the noise spectrum
at the spin Larmor frequency. The ratio of the spectrum amplitudes
at zero- and Larmor frequencies is shown to be a convenient tool
for testing theoretical predictions.
\end{abstract}
\maketitle

Tunneling currents via a microscopic system, such as a quantum
dot, or a molecule or an atom with a localized spin
\cite{bo,ParcolletHooley,KorotkovAverin} attracts considerable
attention in the context of the problem of quantum information
processing. The tunneling current depends on spin-dynamics and
thus encodes its features; at the same time, the tunneling
(measuring) current influences the spin dynamics itself. Thus the
tunneling via a single spin current measurement can provide
information on spin orientation and its dynamics and offer an
example of an indirect-continuous quantum measurement~\cite{bo}.

Scanning tunneling microscopy (STM) experiments \cite{Manassen} on
a single molecule  with a spin in the presence of a magnetic field
${\bf B}_0$, have revealed a peak in the current noise power
spectrum  (i.~e., in the current autocorrelation function) ${\cal
P}(\omega)$  at the Larmor frequency $\omega_L=\gamma B_0$, where
$\gamma$ is the gyromagnetic ratio. Experiments were performed at
room temperatures and found the signal-to-noise ratio $R$ (the
ratio of the power at peak frequency to the shot noise power) to
exceed unity and be almost independent of the orientation of the
applied magnetic field ${\bf B}_0$. Experimental results
\cite{Manassen} are hard to explain within the framework of a
single-spin non-relativistic model  since electrons in the leads
are polarized by the same magnetic field that acts on the spin
i.e. along ${\bf B}_0$, and do not couple with the oscillatory,
components of the spin, which  are perpendicular to ${\bf B}_0$. A
possible relevance of the spin-orbit interaction has been
suggested \cite{Balatsky}. Recently, Levitov and Rashba
\cite{LevitovRashba} noticed that in the systems with the low
space symmetry (such as a dot or a molecule near the surface) the
nonvanishing orbital moment of tunneling electrons couples them to
the mediating spin. They suggested that this mechanism may lead to
a significant effect of the spin oscillatory component on the
tunneling current.

Yet to understand experimental results \cite{Manassen}
 not only the existence of the peak in the
current power spectrum is to be explained, but also the origin of
the large, $R>1$, signal-to-noise ratio and its weak dependence on
the orientation of ${\bf B}_0$. An important step towards
formulation such a model has been done by Bulaevskii, Hru\v{s}ka,
and Ortiz who included in the model the non-relativistic exchange
coupling of a single spin $1/2$ and the tunneling
electrons~\cite{bo,BHO}. Their approach that
followed~\cite{ParcolletHooley} was based on the Keldysh formalism
\cite{Keldysh} and the Majorana-fermion representation
\cite{Tsvelik} for the spin, thus taking into account the
nonequilibrium effects in spin dynamics explicitly.  They found
the spin distribution function and the current-current correlation
function and discussed the dependence of $R$ and line width
$\Gamma$ in the current power spectrum on the applied voltage $V$
between leads, the applied magnetic field ${\bf B}_0$, the
temperature $T$.  They also obtained $R$- and $\Gamma$-dependence
on the degree and orientation ${\bf m}_{\alpha 0}$ of electron
polarization in the right- ($\alpha =$ R) and left ($\alpha =$ L)
current-leads in the steady state (this state establishes during
the transient time after the voltage or tunneling matrix elements
are switched on)~\cite{BHO}. The results of~\cite{BHO} explained
several qualitative features of both average tunneling current
through the spin and noise spectrum at the Larmor frequency. The
quantitative agreement with the experiment was not achieved, yet
the model of ~\cite{BHO} is attractive and warrants to be explored
further. The task now is to identify experimentally accessible
effects that could test the underlying physics of the
tunneling-through-a-spin phenomenon.

We propose measurements of the low-frequency noise (LFN) in the
tunneling current as such a probe. In this Letter we develop a
theory of LFN
 of the tunneling current adapting the Ref.~\onlinecite{BHO} model. We
 predict a tail near zero frequency in the noise spectrum having
 the width of the same order of magnitude as that of the peak at $\omega_L$.
The LFN is expressed through the same quantities as the noise at
the the Larmor frequency, and thus the ratio
$p=\cP(0)/\cP(\omega_L)$ turns out to be a function of the bias
voltage $V$, magnetic field ${\bf B}_0$, as well as of
polarization of the leads and of tunneling coupling. Thus the
experimental study of behavior of the parameter $p$ offers a
unique tool to check on our understanding of tunneling through a
localized spin.

\begin{figure}[h]
\centerline{
\includegraphics[width=5cm]{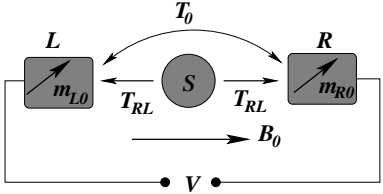}
}
 \caption{ Schematics of the physical system. The electronic
tunneling current is established by a dc  voltage $V$.
\label{fig1}}
\end{figure}

\paragraph {Low-frequency noise of the tunneling current.}
We use the same notations as in~\cite{BHO}: voltage is measured in
the energy units, thus we write just $V$ instead of $eV$,
furthermore $B$ stands for $g\mu_B B$, $T$ stands for $k_B T$, and
$\omega$ represents $\hbar\omega$.  Thus in our notations
$B=\omega_L$. The Hamiltonian is that of the two-leads Kondo
model~\cite{BHO,bo,ParcolletHooley} where the direct tunneling
term is also included:
\begin{eqnarray}
{\cal H}&=&{\cal H}_e+{\cal H}_s+{\cal H}_T, \ \ \
{\cal H}_T={\cal H}_{\sf ref} + {\cal H}_{\sf tr}, \label{Ham} \\
{\cal H}_e&=&\!\!\!\sum_{\alpha, n, \sigma,\sigma'}\!
[\epsilon^{\;}_{n \alpha} \delta_{\sigma\sigma'}-\frac{1}{2}{\bf
B}_{\alpha} \cdot \vec{\bm{\sigma}}_{\sigma\sigma'}]
c^{\dagger}_{\alpha n
\sigma}c^{\;}_{\alpha n \sigma'}, \nonumber \\
{\cal H}_s&=&-g\mu_B {\bf B}_0\cdot{\bf S}\, ,  \nonumber \\
{\cal H}_{\sf ref}&=&\!\!\!\! \sum_{\alpha,n,n' \sigma,
\sigma'}\!\!\!\!\! c^{\dagger}_{\alpha n
\sigma}(\hat{T}_{\alpha\alpha})_{\sigma\sigma'} c^{\;}_{\alpha n'
\sigma' } , \ \ \ \hat{T}_{\alpha\alpha}=T_{\alpha\alpha}^{({\sf
ex})}{\bf S}\cdot\vec{\bm{\sigma}}_{\sigma\sigma'} ,
\nonumber \\
{\cal H}_{{\sf tr}}&=&\!\!\!\!\sum_{n,n', \sigma,\sigma'} \!\!\!\!
c^{\dagger}_{{\rm R} n \sigma} (\hat{T}_{{\rm
RL}})_{\sigma\sigma'} c^{\;}_{{\rm L} n' \sigma' }\! + h. c. ,
\nonumber \\
&&(\hat{T}_{{\rm RL}})_{\sigma\sigma'}= T_0\delta
_{\sigma\sigma'}+ T_{{\rm RL}}^{({\sf ex})}{\bf S} \cdot
\vec{\bm{\sigma}}_{\sigma\sigma'} , \label{t1}
\end{eqnarray}
            where $c^{\dagger}_{\alpha n \sigma}$ ($c^{\;}_{\alpha n
\sigma}$) creates (annihilates) an electron in the left or right
lead (depending on $\alpha \in \{{\rm L,R}\}$) in the eigenstate
$n$, and with spin $\sigma$. Further, $\epsilon^{\;}_{n
\alpha}=\epsilon^{\;}_n -\mu_{\alpha}$,  where $\epsilon_n$ is the
energy in the state $n$ and $\mu_{\alpha}$ is the  chemical
potential in the lead $\alpha$, while $\vec{\bm{\sigma}}$
represents the three Pauli matrices. $T_{{\rm LL}}^{({\sf ex})}$,
$T_{{\rm RR}}^{({\sf ex})}$ and  $T_{{\rm LR}}^{({\sf ex})}$ are
tunneling matrix elements due to the exchange interaction  for the
electron tunneling from the leads to the molecule with the spin
1/2, while  $T_0$ is the direct tunneling matrix element. We take
them as real numbers.  The spin localized in the molecule is
described by the operator ${\bf S}=(S_x,S_y,S_z)$. Figure
\ref{fig1} sketches the physical setup we want to study and which
basically represents the model Hamiltonian $\cal H$.

The electrical current operator can be written as
\begin{equation}
\hat{I}(t)=-ie\!\!\!\sum_{n,n',\sigma,\sigma'} c_{{\rm
R}n\sigma}^{\dagger}(t)(\hat{T}_{{\rm RL}})_{\sigma\sigma'}
c^{\;}_{{\rm L}n'\sigma'}(t) + H.c.\, .
\end{equation}
Since the spin-dependent tunneling amplitude $\hat{T}_{RL}$,
Eq.~(\ref{t1}), contains two terms, the current can be
schematically presented by two vertices: $\hat{T}_0 \equiv T_0\,
\delta_{\sigma \sigma'}$ corresponding to the spin-independent
tunneling, and {$\hat{T}_s \equiv T_{{\rm RL}}^{({\sf ex})}{\bf S}
\cdot \bm{\sigma}_{\sigma \sigma'}$ corresponding to the
spin-dependent part.
\begin{figure}[ht]
\centerline{
\includegraphics[width=8cm]{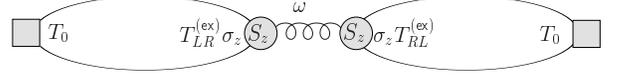}
 } \caption{Skeleton diagrams for calculation of
the low-frequency noise. The wavy line corresponds to the
correlation function $\langle \delta S_z^2 \rangle_\omega$.
\label{fig2}}
\end{figure}

In the following we assume $T_{{\rm RL}}^{({\sf ex})} \ll T_0$.
Since we are interested only in low frequencies, $\omega \ll B,
V$, only $z$-component should be kept. Thus the current noise can
be expressed by the schematic diagram shown in Fig.~\ref{fig2}.
The left and right electron blocks are nothing but the derivatives
$\partial I/\partial \av{S_z}$, while the wavy line corresponds to
the correlation function
$\langle \delta S_z^2 \rangle_\omega \equiv \av{[\delta
S_z(t),\delta S_z(0)]_+}_\omega$.
This way we arrive at the natural expression for the low-frequency
noise,
\begin{equation} \label{eq:002}
{\cal P}(\omega)= \av{[\delta I(t),\delta I(0)]_+}_\omega
=\left(\partial I/\partial \av{S_z} \right)^2 \langle \delta S_z^2
\rangle_\omega\, .
\end{equation}

To be concrete let us  restrict ourselves to the case of fully
polarized electrons in the leads. Then there exists the average
current proportional to the average spin, $\langle \bf S \rangle$,
and given by the expression~\cite{BHO}
\begin{eqnarray}
I(V) &=&
I_0(V)+{\bf I}_s(V)\cdot \langle{\bf S}\rangle, \label{avcur} \\
I_0(V)&=&\pi e (1+{\bf m}_{\rm R}\cdot{\bf m}_{\rm L}) \
T_0^2\rho_0^2 \ V, \label{avcur1}\\
{\bf I}_s(V)&=&2\pi e ({\bf m}_{{\rm R}}+{\bf m}_{{\rm L}}) \
T_0T_{{\rm RL}}^{({\sf ex})} \rho_0^2 \ V. \label{avcur2}
\end{eqnarray}
In the above equations $\rho_0$ is the density of states per spin
(DOS) of the leads at the Fermi level (when leads are different
$\rho_0^2=\rho_0^{\rm L}\rho_0^{\rm R}$ where $\rho_0^\alpha$ is
the DOS in the lead $\alpha$); ${\bf m}_\alpha$ means the
direction of electron spin polarization in the coordinate system
with $z$-axis parallel to the total magnetic field, ${\bf B}={\bf
B}_0+{\bf B}_T$. Here ${\bf B}_0$ is the external magnetic field
while ${\bf B}_T$ is the additional dc magnetic field produced by
tunneling electrons.

The derivative $\partial I/\partial \av{S_z} \equiv I_{sz}(V)$,
see Eq.~(\ref{avcur}), may be found in~\cite{BHO}, and we turn to
calculation of the spin correlation function. In the equilibrium,
the fluctuation-dissipation theorem yields:
\begin{equation} \label{eq:003}
\av{\delta S_z^2}_\omega=\coth \frac{\omega }{2T}\, \Im
\chi(\omega)\approx  \frac{2 \Gamma_z\, \left(
1-4\av{S_z}_T^2\right)}{\omega^2
  +\Gamma_z^2}\, .
\end{equation}
Here $\av{S_z}_T$ is the equilibrium average spin,  $\Gamma_z$ is
the decay rate for the $S_z$ fluctuations.

Far from the equilibrium the fluctuation-dissipation theorem
generally speaking does not apply. However, in a non-equilibrium,
but yet \emph{stationary} state one can still use the result of
formula (\ref{eq:003}) with the appropriate expression for the
average spin, $ \av{S_z}=(1/2)h_f(b)$.
\begin{equation}
\label{eq:006}
 \av{\delta S_z^2}_\omega =  \frac{2
 \Gamma_z \left[1-h_f^2(v,b)\right]}{\omega^2 +\Gamma_z^2}\, , \quad b=\frac{B}{T}, \ v=\frac{V}{T}\, .
\end{equation}
 Here $h_f (b,v)$ is the function calculated
in~\cite{BHO}:
\begin{eqnarray}
h_{f}\!\!\!&=& \!\!\!-\frac{2{ b}(1-m_{{\rm R} z} m_{{\rm L} z})
\!-\! 2 {v} ( m_{{\rm R} z} - m_{{\rm L} z}) \! + \!{ b} \
\theta_1}{\phi^+  (1-m_{{\rm R} z} m_{{\rm L} z})\!-\!  \phi^-
(m_{{\rm R} z} - m_{{\rm L} z})\! + \!\phi ({ b}) \ \theta_1} ,
\nonumber \\
\theta_1 &=& \frac{T_{{\rm RR}}^2 ( 1-m_{{\rm R }z} ^2) + T_{{\rm
LL}}^2
( 1-m_{{\rm L}z} ^2)}{T_{{\rm RL}}^2} , \quad v=\frac{V}{T}, \nonumber \\
\phi(b)&=&b\coth(b/2)\, , \quad \phi^{\pm} \equiv \phi({v}+{b})\pm
\phi({v}-{b}) \,. \label{hBgenP}
\end{eqnarray}
 Equation (\ref{hBgenP}) makes sense
provided  $m_{{\rm R} z}\neq 1$ and $m_{{\rm L} z} \neq \pm 1$.
Otherwise, if $m_{{\rm R} z} = 1$, $m_{{\rm L} z} = \pm 1$, the
self-consistency equation is an identity and the spin steady state
can be any.

Equation (\ref{eq:006}) can be derived, e.~g., using the the
technique developed by Abrikosov~\cite{abric}, where a 1/2-spin
was interpreted as a pseudo-Fermion with the Green function
$g_\pm^{(0)}(\epsilon)=(\epsilon \mp B/2- \lambda + i
\delta)^{-1}$.
Here $\lambda$ is an auxiliary ``chemical potential'' which is
send to infinity eventually. This trick allows one to remove extra
unphysical states that appear because Fermi operators have more
extended phase space than spin operators. The method was
elaborated by Maleev for the case of dynamical defects in
glasses~\cite{maleev}. Schematic diagrammatic representation of
the correlation function $\av{\delta S^2}_\omega$ is given in
Fig.~\ref{fig:diag}. Derivations can be carried out similarly to
those in~\cite{Bergli} where electron dephasing rates due to
pseudo-spin defects were calculated. A similar procedure was also
used for calculation of the energy relaxation time of the
electrons in a thin wire due to magnetic
impurities~\cite{Goppert}.
\begin{figure}[b]
\includegraphics[width=8cm]{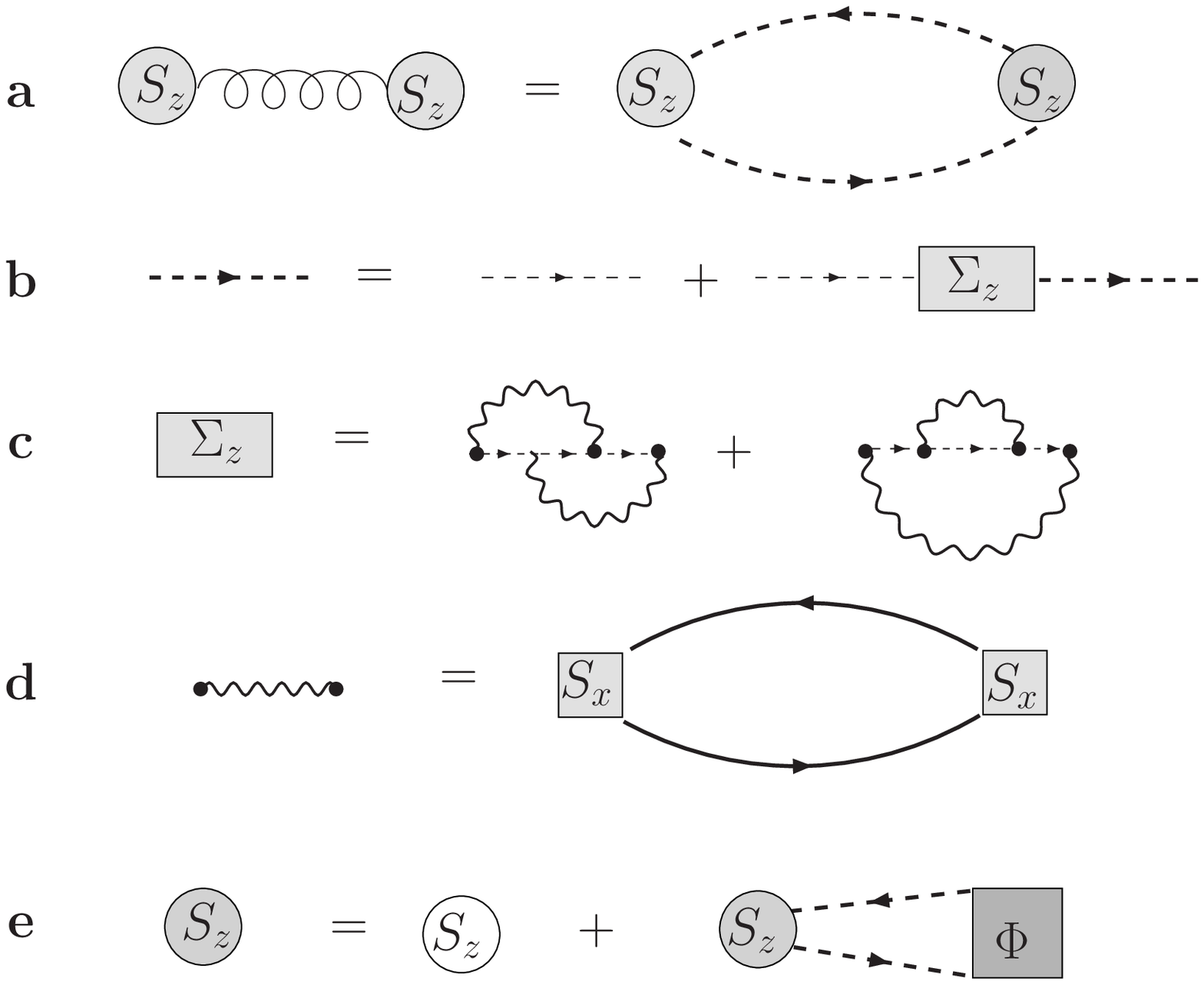} \\ \vspace*{0.2in}
\hspace*{-1cm}\includegraphics[width=7cm]{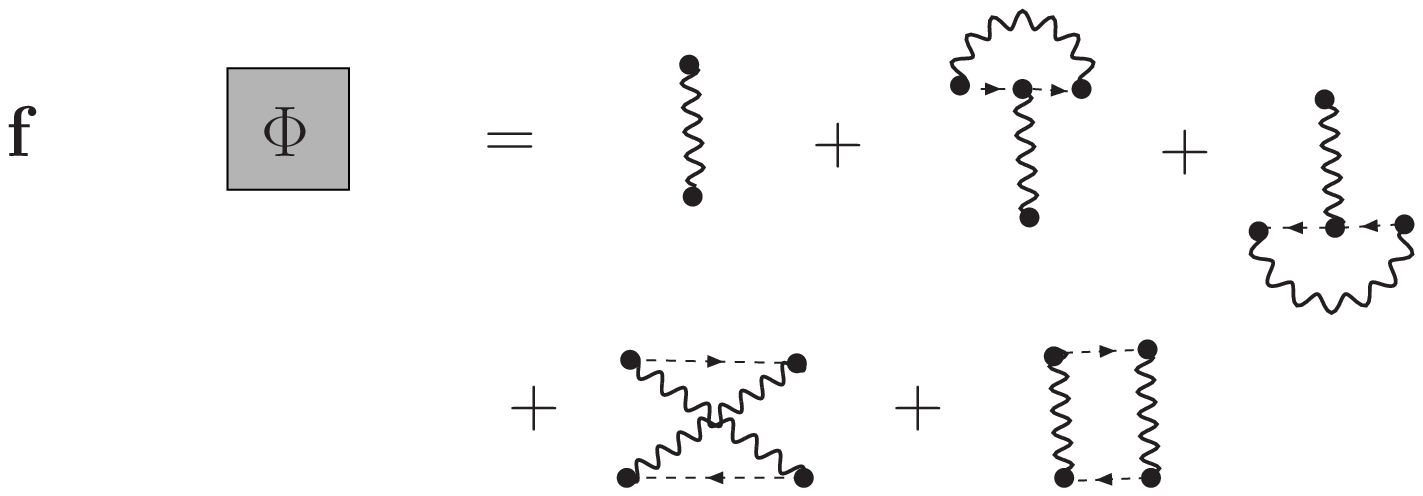}
 \caption{Schematic diagrams for calculation of $\av{\delta
     S^2}_\omega$. Dashed lines represent auxiliary fermions, while
     the vertices $S_z$ and $S_x$ represent the tunneling amplitudes
     $T^{\sf (ex)}_{RL}\sigma_z$ and $T^{\sf (ex)}_{RL}\sigma_x$, respectively. \label{fig:diag}}
\end{figure}
The correlation function is given by the diagram shown in
Fig.~\ref{fig:diag}a where thick dashed line corresponds to the
auxiliary fermion $g_\pm (\epsilon)$.  An important feature of the
calculation is that for calculation of $\av{\delta S_z^2}_\omega$
one has to calculate the self energy up to the \textit{fourth}
order in $T_{RL}^{\sf (ex)}\sigma_x$ as shown in
Figs.~\ref{fig:diag}b and \ref{fig:diag}c , see also Appendix
in~\cite{maleev}. The wavy line, Fig.~\ref{fig:diag}d, corresponds
to $S_x-S_x$ propagator and solid lines represent tunneling
electrons. This propagator was actually calculated in~\cite{BHO}.
A similar complication arises when calculating the $S_z$ (namely,
$T_{RL}^{\sf{(ex}}\sigma_zS_z$) vertex part, see
Fig.~\ref{fig:diag}e,f.
When ${\bf m}_{{\rm R}}={\bf m}_{{\rm L}}$ (except for $m_{\alpha
z}=\pm 1$), the  result is the same as for unpolarized
electrons~\cite{ParcolletHooley}:
\begin{equation}
h_f(b,v)=\tanh\left(\frac{b}{2}\right) \
\frac{\phi({b})(2+\theta)} {\phi^++\phi({ b})\theta}\, .
\end{equation}
Here $\theta=(T_{{\rm LL}}^2+T_{{\rm RR}}^2)/T_{{\rm RL}}^2$. We
would like to emphasize that matrix elements $T_{\alpha \alpha}$
describe electron tunneling from the lead $\alpha$ to the spin
site and back while $T_{LR}$ describes spin mediated tunneling
{\it between} the leads. Thus the value of $\theta$ is extremely
sensitive to the location of the spin with respect to the leads.
For the perfectly symmetric configuration $T_{LL} = T_{RR} =
T_{LR}$, therefore $\theta = 1/2$. Since both $T_{LL}$ and
$T_{RR}$ increase dramatically  with a decrease of the distance
from the spin to the corresponding lead so does $\theta$ with an
increase in asymmetry. As a result, in the asymmetric
configurations the average spin is actually equal to its
equilibrium value. This was noticed, in particular, for the
similar problem of the electron tunneling mediated by the presence
of the structural two-level system in \cite{kozub}. However for
symmetric configuration the spin coupling to the electrons
tunneling between the leads is as strong as its coupling to the
electrons in any of the leads. As a result, the average spin is
controlled by the combination of electron energy distributions
within both of the leads, and it is out of equilibrium provided $v
> 1$.  The tunneling electrons reduce the spin magnetization which
drops as $1/V$ at large $V$.

Substituting Eq.~(\ref{eq:006}) into Eq.~(\ref{eq:002}) we obtain
\begin{equation} \label{eq:noi001}
\cP_\omega= \frac{2
  \Gamma_z}{\omega^2
  +\Gamma_z^2}\cdot \, \left(\frac{\partial I}{\partial \av{S_z}
  }\right)^2\left[1-h_f^2(v,b) \right]\, .
\end{equation}
Using the expression from~\cite{BHO} for the spin-dependent part
of the current, $$\partial I/\partial \langle{\bf S}\rangle \equiv
{\bf I}_s(V)= 2\pi e ({\bf m}_R +{\bf
m}_L)T_0T_{RL}^{\text{(ex)}}\rho_0^2V \, , $$ we arrive at the
final expression for the low-frequency noise:
\begin{eqnarray} \label{eq:noif}
\cP_\omega &=& P_0\, v^2 \left[1-h_f^2(v,b) \right]
  \, \frac{2
  \Gamma_z}{\omega^2
  +\Gamma_z^2}\, , \\
P_0&=&\left[2\pi T_0T_{RL}^{\text{(ex)}}\rho_0^2 T\right]^2\!
(m_{Rz} + m_{Lz})^2\, . \label{eq:pol}
\end{eqnarray}
This result agrees with calculations by Shnirman \textit{et
al.}~\cite{smm}. According to Eq.~(\ref{eq:pol}), the effect is
strongly dependent on a degree of electron spin polarization. Here
we note that the polarization can arise not only due to
spin-dependent tunneling amplitude, but also as a result of
electron motion through the molecule where the localized spin is
located. Indeed, as it is known for semiconductor
structures~\cite{Perel}, the electrons tunneling through the
barriers with no inversion center become spin-polarized if the
tunneling electron has a component of wave vector $k_{\parallel}$
parallel to the barrier plane. The difference between the
tunneling exponents for opposite spin direction can be estimated
as $ \sim \gamma (2mk_{\parallel}/\hbar^2)kd $ where $\gamma$
characterizes an efficiency of spin-orbital interaction (which for
typical semiconductors is of the order of $10^{-36}$~erg$\cdot$
cm$^3$) while $d$ is the tunneling length. Of course, there is a
difference between the organic molecule and a semiconductor.
However the apparent absence of any pronounced symmetry in the
considered  situation  can result in a finite $k_{\parallel}$,
which  will, in turn, lead to a spin polarization. In particular,
for $k_{\parallel} \simeq 10^6$ cm$^{-1}$ (two orders of magnitude
less than typical atomic value) and for the values of $\gamma$ two
order of magnitude less than the estimate given above the degree
of spin polarization can be of the order of $10^{-4}$ which is
larger than the spin polarization of electrons in metals at $B
\sim 1$ T. Thus the factor mentioned above can significantly
increase the coupling of the tunneling electrons with localized
spin.

\paragraph{Comparison with the noise at Larmor frequency.} According to
Eq.~(\ref{eq:noif}),  the low-frequency noise is represented by a
Lorentzian tail with the width $\Gamma_z$. At $\omega \to 0$
\begin{equation} \label{eq:noiz}
\cP_0 \equiv \cP_\omega \vert_{\omega=0}=2P_0\, v^2
\left[1-h_f^2(v,b)\right]\Gamma_z^{-1}\, .
\end{equation}
 Let us compare this result with the maximal value of the noise at Larmor frequency,
 $\cP_L$. 
 According to~\cite{BHO}, the magnitude of the noise near the Larmor frequency
at $V>B$ is given by the expression
\begin{eqnarray}
&&{\cal P}_\omega= P_1 [v^2+v b \, h_f(v,b)]\,
\frac{\Gamma_{\perp}}
{\Gamma_{\perp}^2+(\omega-B)^2}, \label{gr} \\
&&P_1=(\pi T_0T_{\rm RL}^{\text{ex}}\rho_0^2T)^2 |{\bf m}_{R
\perp} +{\bf m}_{L \perp}|^2 \nonumber
\end{eqnarray}
(for fully polarized electrons). At $V <B$, ${\cal P}_\omega=0$.

Comparing now noise magnitudes at $\omega \to 0$ and at $\omega
\to \omega_L$ respectively at $V>B$, we have
\begin{eqnarray}
&&\cP_L  \equiv  \cP_\omega
\vert_{\omega=\omega_L}=(P_1/\Gamma_\perp)\, [v^2+v b \,
h_f(v,b)]\, , \label{eq:noi2} \\
&&  p \equiv \frac{\cP_0 }{\cP_L}=8
\frac{\Gamma_\perp}{\Gamma_z}\,\left|\frac{m_{Rz} + m_{Lz}}{{\bf
m}_{R
\perp} +{\bf m}_{L \perp}}\right|^2{\cal F}(v,b)\, ,  \\
 &&{\cal F}(v,b)
=[1-h_f^2(v,b)]/[1+(b/v)\,h_f(v,b)]\, . \label{eq:noi3}
\end{eqnarray}
It rather difficult to provide realistic estimates for the ratio
$\Gamma_\perp / \Gamma_z$ since spin relaxation and dephasing can
be produced both by tunneling electrons and by some degrees of
freedom in the leads. The contributions of the tunneling electrons
to $\Gamma_\perp$ and $\Gamma_z$ are calculated in the
Bloch-Redfield approximation in Ref.~\onlinecite{smm}. What is
important that in general
 $\Gamma_\perp \gtrsim \Gamma_z/2$, and the ratio $\Gamma_\perp / \Gamma_z$
 can be measured experimentally. The dimensionless function ${\cal
 F}(v,b)$ grows monotonously with $v$ for a given $b$, it tends to 1 at $v \gg b$,
 the plot ${\cal F}(v)$ shifts slightly downwards as parameter $b$
increases. Thus, the ratio $p$ can be of the order of unity.

In conclusion, we have calculated the low-frequency noise power in
the tunneling current, ${\cal P}(0)$, and demonstrated that the
ratio $p(V,T)={\cal P}(0)/{\cal P}(\omega_L)$ is the universal
function of dimensionless voltage and magnetic field, $eV/k_BT$
and $g\mu_B B/k_BT$, respectively. This opens the route for
identification underlying mechanisms of the noise in tunneling
current by comparison of the measured dependence of $p(V,T)$ upon
voltage and temperature with the obtained ${\cal F}(V,T)$. We also
noted that spin polarization can be obtained due to intrinsic spin
polarization while tunneling through a complex molecule.

 This research is supported by the US DOE Office of
Science under contract No. W-31-109-ENG-38. We like to thank
L.~Bulaevskii and M.~Hru\v{s}ka for useful discussions. We are
specifically thankful to A. Shnirman who pointed out some mistakes
in our original draft and made Ref.~\onlinecite{smm} available.

\end{document}